\documentclass[apj]{emulateapj}
\usepackage{apjfonts}
\usepackage{color} 
\pdfoutput=1

\newcommand{\src}{GS~1826$-$24}
\newcommand{\fluxcgs}{erg cm$^{-2}$ s$^{-1}$}

\begin{document}
 
\title{Constraints on neutron star mass and radius in \src\ from sub-Eddington X-ray bursts}
 
\author{Michael Zamfir\altaffilmark{1}, Andrew Cumming\altaffilmark{1}, and Duncan K. Galloway\altaffilmark{2}}

\altaffiltext{1}{Department of Physics, McGill University, 3600 rue University, Montreal, QC, H3A 2T8, Canada; mzamfir@physics.mcgill.ca, cumming@physics.mcgill.ca}
\altaffiltext{2}{Monash Centre for Astrophysics (MoCA), School of Physics, and School of Mathematical Sciences, Monash University, VIC 3800, Australia; Duncan.Galloway@monash.edu}

\begin{abstract}
We investigate the constraints on neutron star mass and radius in \src\ from models of lightcurves and spectral evolution of type I X-ray bursts. This source shows remarkable agreement with theoretical calculations of burst energies, recurrence times, and lightcurves. We first exploit this agreement to set the overall luminosity scale of the observed bursts. When combined with a measured blackbody normalization, this leads to a distance and anisotropy independent measurement of the ratio between the redshift $1+z$ and color correction factor $f_c$. We find $1+z=1.19$--$1.28$ for $f_c=1.4$--$1.5$. We then compare the evolution of the blackbody normalization with flux in the cooling tail of bursts with predictions from spectral models of Suleimanov et al.~(2011b). The observations are well described by the models at luminosities greater than about one third of the peak luminosity, with deviations emerging at luminosities below that. We show that this comparison leads to distance independent upper limits on $R_\infty$ and neutron star mass of $R_\infty\lesssim 9.0$--$13.2\ {\rm km}$ and $M<1.2$--$1.7\ M_\odot$, respectively, for solar abundance of hydrogen at the photosphere and a range of metallicity and surface gravity. The radius limits are low in comparison to previous measurements. This may be indicative of a subsolar hydrogen fraction in the \src\ photosphere, or of larger color corrections than that predicted by spetral models. Our analysis also gives an upper limit on the distance to \src\ of $d<4.0$--$5.5\ {\rm kpc}\ \xi_b^{-1/2}$, where $\xi_b$ is the degree of anisotropy of the burst emission.
\end{abstract}


\section{Introduction}

Thermonuclear flashes from accreting neutron stars, observed as type I X-ray bursts, in principle provide a way to constrain neutron star masses and radii (for a review see \S 4 of Lewin, van Paradijs \& Taam 1993). The large observational catalogues of type I X-ray bursts now available (Galloway et al.~2008; hereafter G08) and new spectral models (Madej, Joss, \& Rozanska 2004; Majczyna et al.~2005; Suleimanov, Poutanen, \& Werner 2011b) have motivated fresh attempts to do this using photospheric radius expansion bursts (\"Ozel 2006; {\"O}zel, G{\"u}ver, \& Psaltis 2009; G\"uver et al.~2010a, 2010b; Steiner et al.~2010; Suleimanov et al.~2011a, 2011b; \"Ozel, Gould, \& G\"uver 2011). In this approach, the peak luminosity of the burst (specifically at the ``touchdown'' point when the photosphere returns to the neutron star surface) is related to the Eddington luminosity, and the normalization of the burst spectrum is related to the emitting area. If some information about the distance to the source is available, constraints on the neutron star mass and radius can be derived. These works have highlighted and spurred debate about some of the systematic errors that must be taken into account, such as uncertainty in identifying the moment at which the photosphere touches down (Galloway, \"Ozel, \& Psaltis 2008; Steiner et al.~2010; Suleimanov et al.~2011a,b; G\"uver, \"Ozel \& Psaltis 2011), and differences in derived radii when using bursts at different accretion rates from the same source (Suleimanov et al.~2011a,b; G\"uver, Psaltis, \& \"Ozel 2011).

\src\ is a unique X-ray burster that shows remarkable agreement with theoretical models of recurrence times, energetics, and lightcurves (Galloway et al.~2004; Heger et al.~2007; in 't Zand et al.~2009). The observed recurrence time (typically 3--5 hours) in a given epoch is the same from burst to burst to within a few minutes (Cocchi et al.~2001), and the burst lightcurves in a given epoch are very uniform (Galloway et al.~2004), implying the same conditions on the neutron star surface from burst to burst, and a regular limit cycle. Heger et al.~(2007) compared the observed lightcurves to the theoretical models of Woosley et al.~(2004). By choosing a model with approximately the same recurrence time as the data, and by varying the distance only (which scales the observed peak flux to match the model peak luminosity), the theoretical lightcurve fit most of the observed lightcurve well, except for deviations during the burst rise and at late times in the cooling tail. This is of great interest because the long $\sim 100$ second tails of these bursts are powered by the rp-process (Wallace \& Woosley 1981), and offer a way to test the nuclear physics input, such as masses of proton rich heavy nuclei and their reaction rates, both of which are usually highly uncertain (e.g.~Schatz et al.~1998; Schatz 2006). Even more remarkably, in 't Zand et al.~(2009) extracted the observed lightcurve out to more than $1000$ seconds by combining multiple bursts. The late time cooling observed matched the late time cooling in the theoretical model of Heger et al.~(2007), which arises from heat initially conducted inwards to deeper layers that then emerges on long timescales.

In this paper, we take the comparison between observations and theory for \src\ one step further. We first determine the constraints on neutron star mass and radius that can be derived from the lightcurve comparison carried out by Heger et al.~(2007). The basic idea here is that even though the X-ray bursts from \src\ do not reach the Eddington limit, we can still determine the intrinsic luminosity of the bursts by comparing with the lightcurve models. This replaces the touchdown measurement used for PRE bursts with a different condition which we use to constrain $M$ and $R$ for \src. In our analysis we take care to include the possible anisotropy in burst and persistent emission and show how that could affect the mass and radius determination. We then compare the spectral evolution during the tail of the burst with spectral models. The good understanding of bursts from \src\ suggests that they could be a good testing ground for spectral models. We show that in the initial cooling phase following peak luminosity the spectral evolution agrees well with the models of Suleimanov et al.~(2011b), and we derive the associated constraints on $M$ and $R$. In both cases, we look for constraints that are independent of distance and emission anisotropies since neither are well-constrained for \src. 

The outline of the paper is as follows. The data analysis is described in \S 2. In \S 3, we discuss the possible anisotropy of the burst and persistent emission and review calculations of the expected degree of anisotropy in the literature. In \S 4, we use the model lightcurve from Heger et al.~(2007) to set the luminosity scale of the observed bursts and show that this gives a distance-independent relation between the redshift and color correction factor $f_c$. Suleimanov et al.~(2011a) argued that rather than using a single measurement of touchdown flux, the entire cooling track of the burst should be fit to spectral models. We do this in \S 5, and show that even though the peak flux is below Eddington, the fits provide a constraint on the value of $F_{\rm Edd}$ as well as the normalization of the spectrum. These two measurements translate into a distance independent upper limit on $R_\infty$. We compare these two different constraints and discuss their implications in \S 6.

\section{Data Analysis}

We used data taken with the Proportional Counter Array (PCA; Jahoda et al. 1996) onboard the Rossi X-ray Timing Explorer (RXTE), from the catalogue of bursts detected over the mission lifetime (G08). Where not explicitly stated, the data analysis procedures are as in G08. Time-resolved spectra in the range 2-60 keV covering the burst duration were extracted on intervals as short as 0.25 s during the burst rise and peak, with the bin size increasing step-wise into the burst tail to maintain roughly the same signal-to-noise level. A spectrum taken from a 16-s interval prior to the burst was adopted as the background.

We re-fit the spectra over the energy range 2.5-20 keV using the revised PCA response matrices, v11.7\footnote{see http://www.universe.nasa.gov/xrays/programs/rxte/pca/doc/rmf/pcarmf-11.7}, and adopted the recommended systematic error of 0.5\%. The fitting was undertaken using XSPEC version 12. In order to accommodate spectral bins with low count rates, we adopted Churazov weighting. We modelled the effects of interstellar absorption, using a multiplicative model component ({\tt wabs} in {\sc XSpec}), with the column density $N_H$ frozen at $4\times10^{21}\ {\rm cm^{-2}}$ (e.g. in 't Zand et al. 1999). In the original analysis carried out by G08, the neutral absorption was determined separately for each burst, from the mean value obtained for spectral fits carried out with the $N_H$ value free to vary. This has a negligible effect on the fluxes, but can introduce spurious burst-to-burst variations in the blackbody normalisation.

The burst data used here has been corrected for ``deadtime'', a short period of inactivity in the detectors following the detection of a X-ray photon. There are however concerns regarding the absolute flux calibration of the PCA associated with variations in the flux from the Crab nebula and the effective area of the PCA. We will show that such absolute uncertainties will not influence our derived constraints.

\section{Anisotropy of the burst and persistent emission}

The possibility that the burst or persistent emission is not isotropic has been long discussed (e.g. Lapidus et al.~1985), but has not always been included in recent work using X-ray bursts to constrain neutron star mass and radius. For example, in \"Ozel (2006), Steiner et al.~(2010), and Suleimanov et al.~(2011a) the burst emission is assumed to be isotropic. Here, we review the expected size of the anisotropy. We follow Fujimoto (1988) and define an anisotropy parameter $\xi$ by the relation $4\pi d^2 F \xi = L$ between the observed flux $F$ and the luminosity of the source $L$ over the whole sky, where $d$ is the distance to the source. When $\xi<1$ ($>1$), the radiation is beamed towards (away from) the observer. We write the anisotropy factor for the burst and persistent emission as $\xi_b$ and $\xi_p$ respectively.

Lapidus, Sunyaev \& Titarchuk (1985) showed that if the accretion disk extends to the neutron star surface during the flash, it will intercept $\approx 1/4$ of the radiation from the burst, reflecting it preferentially along the disk axis. They provide the approximate expression 
\begin{equation}\label{eq:xib}
\xi_b^{-1}={1\over 2}+\left|\cos i\right|
\end{equation}
where $i$ is the inclination angle ($i=0^\circ$ means the system is viewed face on, looking down the disk axis), which closely fits their more detailed results derived from solving the radiative transfer equations for a disk geometry (they found a maximum value of 1.39 rather than 1.5). The range of $\xi_b^{-1}$ is from 0.5 (edge on) to 1.5 (face on), implying an uncertainty of a factor of 3 depending on inclination angle.

The anisotropy factor for the persistent emission is perhaps even more uncertain than that for the burst flux, depending on the specific model for the inner accretion disk, boundary layer, and corona etc. Lapidus et al. (1985) and Fujimoto (1988) derive opposite behaviors for the factor $\xi_p$ as a function of inclination. The model presented in Fujimoto (1988) for the persistent emission assumes that radiation from the boundary layer, which encircles the neutron star in a ``belt'' about its equator, is largely screened by the inflated inner part of the accretion disk and scattered preferentially in a direction along the disk axis. In Lapidus et al.~(1985), however, the inner part of the disk is assumed to be thin, and less than one half of the boundary layer radiation falls on the accretion disk and is re-scattered, again preferentially along the disk axis, while the remainder of the emission is beamed preferentially in the direction $i=90^\circ$ (along the plane of the disk). This difference in their modelling of the inner accretion disk is made apparent by fact that, while Fujimoto (1988) predicts no radiation to be emitted in the $i=90^\circ$ direction, Lapidus et al.~(1985) find a substantial portion of the persistent emission will be beamed in that direction. The ratio $\xi_p/\xi_b$ varies by up to a factor of $\sim$3 with inclination for both models, although while Fujimoto (1988) finds that the ratio is monotonically increasing with inclination, Lapidus et al.~(1985) find the opposite trend. We note that these two models do not consider the effects of general relativity on the trajectories of photons near the neutron star surface. However, Lapidus et al.~(1985) show that when considering this effect, a substantially larger proportion ($\sim28\%$ for a ratio of the neutron star radius to the Schwarzschild radius $R/r_{s}$ of 3) of radiation falls on the accretion disk, further enhancing beaming along the disk axis.

The definition of $\xi$ is such that it always appears with distance $d$ in the combination $\xi^{1/2}d$. Therefore, the uncertainty in anisotropy factor acts in the same way as an additional uncertainty in the distance to the source. For \src, Homer et al. (1998) suggest a limit $i<70^\circ$ based on the low amplitude of the optical modulation at the orbital frequency, in which case equation (\ref{eq:xib}) gives $0.84\lesssim \xi_b^{-1}\lesssim 1.5$, or $0.85<\xi_b^{1/2}<1.1$. Therefore even if the distance to \src\ was perfectly known, anisotropy of the burst emission would represent an uncertainty of about $\pm 15$\% in any quantity that depends on distance. For example, using spectral fits to determine $R_\infty$ is subject to this uncertainty since the normalization of the spectrum depends on the solid angle $R^2_\infty/d^2\xi$. Given these uncertainties, in this paper we look for constraints on $M$ and $R$ that are independent of distance and anisotropy.

\section{Comparison between the observed and model burst lightcurves}

We first ask what constraints on $M$ and $R$ arise from the comparison between the observed lightcurve and theoretical models of Heger et al.~(2007). Photospheric radius expansion (PRE) bursts are often used in work to constrain neutron star properties from X-ray bursts, because the peak luminosity of the burst can then be taken to be the Eddington luminosity. This cannot be done for \src\ because the bursts do not show PRE, implying that they have a peak luminosity below Eddington. Instead, here we pursue the idea that the model lightcurves which fit the observed lightcurves so well tell us the peak luminosity of the bursts.

Heger et al.~(2007) selected from their models one that had a similar recurrence time to the observed bursts in 2000 (the model had $t_{\rm recur}=3.9$ hr as opposed to the observed $t_{\rm recur}=4.07$ hr). They showed that, when the distance to \src\ (actually $\xi_b^{1/2}d$) is chosen to make the predicted peak flux match the observed lightcurve, the theoretical and observed burst lightcurves show remarkable agreement. They considered fixed values of $M$ and $R$, but the choice of those two parameters also changes the mapping between observed and model burst fluxes. Therefore, rather than vary distance alone, we find the value of the ratio of the observed flux to the model flux 
\begin{equation}\label{eq:Fratio}
{F_{\rm obs}\over F_{\rm model}}=\xi_b^{-1}\left({R\over d}\right)^2 {1\over \left(1+z\right)^2}
\end{equation}
that gives the best fit between the model and the data. Taking the peak values, $F_{\rm model, pk}=1.29\times 10^{25}$ \fluxcgs$\,$ (computed from the redshifted peak luminosity quoted in Heger et al.~2007, with $R=11.2\,$km and $z=0.26$) and $F_{\rm obs,pk}=2.84\times 10^{-8}$ \fluxcgs, we find $F_{\rm obs}/F_{\rm model}=2.20\times 10^{-33}$. Substituting this value into equation (\ref{eq:Fratio}), we find
\begin{equation}\label{eq:Fratio2}
{R\over \xi_b^{1/2}d}=10.9\ {\rm km/6.0\ {\rm kpc}}\ {1+z\over 1.26} \left({F_{\rm obs}/F_{\rm model}\over 2.2\times 10^{-33}}\right)^{1/2},
\end{equation}
where we use the redshift assumed by Heger et al.~(2007). Note that the model lightcurve is likely to be insensitive to the model gravity, so that the ratio $F_{\rm obs}/F_{\rm model}$ does not depend sensitively on the $M$ and $R$ used in the model. For example,  the ignition column depth is weakly dependent on gravity in this burning regime (Bildsten 1998 derives $y_{\rm ign}\propto g^{-2/9}$). However, this is something that should be explored in further simulations. For now, we assume $F_{\rm obs}/F_{\rm model}$ is a constant, and take equation (\ref{eq:Fratio2}) as a joint constraint on $R$ and $1+z$.

The theoretical uncertainty in $F_{\rm model}$ is at present unknown. The predicted lightcurves depend on the input nuclear physics, and prescription for convection and other mixing processes for example. These prescriptions vary from code to code, and currently only simulations from the Kepler code (Woosley et al.~2004) have been compared to the observations of \src.  Further simulations and comparisons are required to determine what range of predicted peak fluxes still produce lightcurves with the correct shape to fit the data. For now, in order to put an error bar on the prefactor in equation (\ref{eq:Fratio2}), we assume that the theoretical uncertainty in $F_{\rm obs}/F_{\rm model}$ is $\pm 10$\%, and keep in mind the fact that this number is uncertain.

\begin{figure}
\includegraphics[width=\columnwidth,height=68mm]{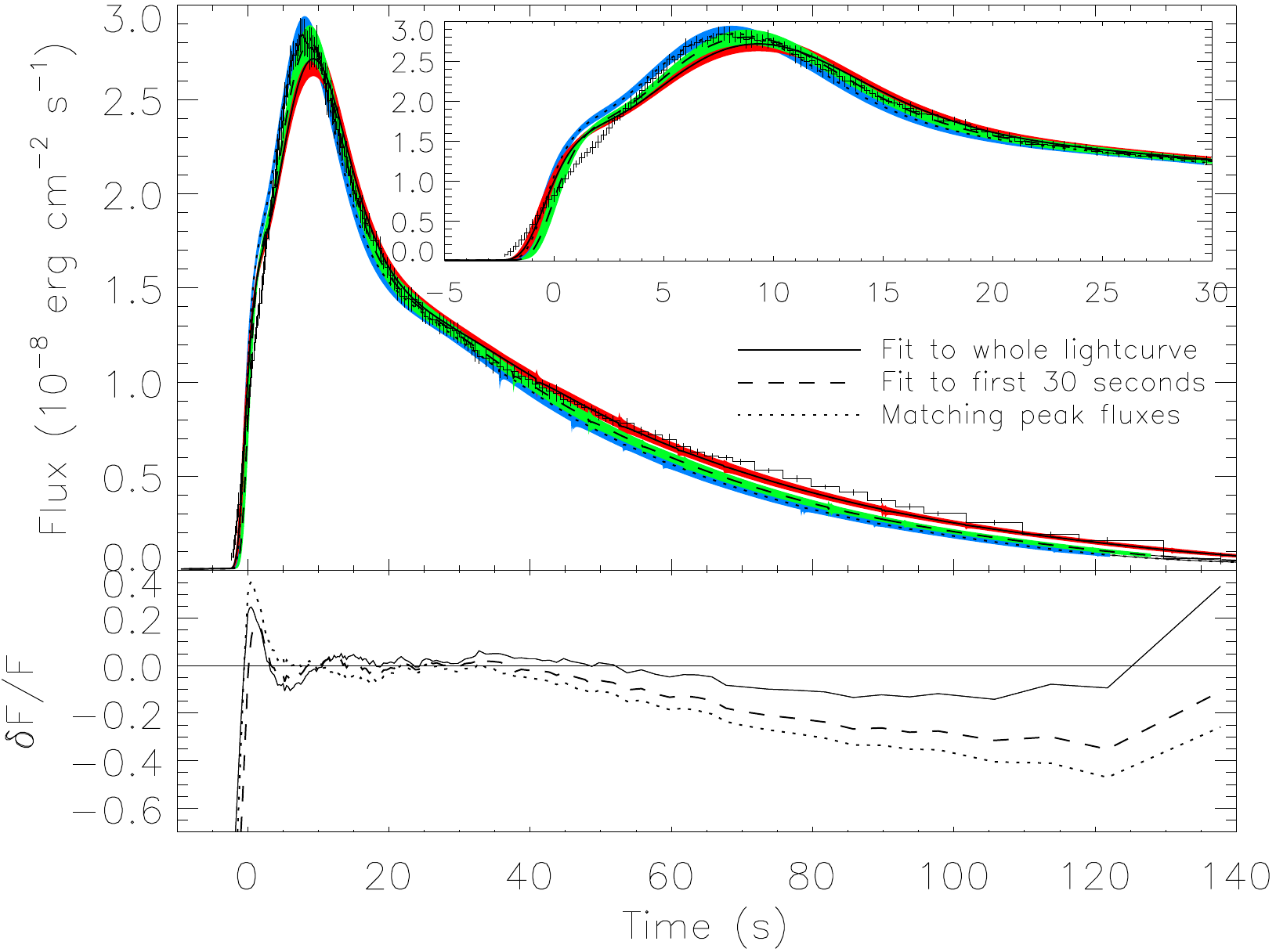}
\caption{The average burst profiles with $t_{\rm recur}=4.07$ hr compared with three separate fits of the mean theoretical model lightcurve from Heger et al. (2007) (model A3 which had a similar recurrence time) is shown in the upper plot, with an inset showing only the first 30 seconds. The model has been fit to the data by varying the overall normalization, start time, and redshift. The solid (with red band) and dashed (with green band) lines represent the fits to the entire lightcurve and the first 30 seconds, respectively, and the dotted line (with blue band) represents the lightcurve fitted only by matching the peak fluxes with a fixed redshift of $1+z=1.26$. The bands show the range of luminosity variations from burst to burst in the theoretical model. The lower plot shows the difference between each of the fits and the average observed lightcurve, with a horizontal solid line at $\delta F/F=0$ for clarity.}
\label{fig:lc}
\end{figure}

This raises the point that rather than use the peak flux only, we could also fit the entire lightcurve. In that case there is an extra parameter, the redshift $1+z$ which stretches the lightcurve in time. In principle, this provides a constraint on $1+z$. In practice, however, we find that the value of $1+z$ obtained in the fit is sensitive to how much of the lightcurve is included in the fit. For example, fitting the entire lightcurve (until about 130s after the peak) we find best-fit values $1+z=1.44$, $F_{\rm obs}/F_{\rm model}=2.10\times 10^{-33}$. If we fit the first 30 seconds only, which includes only the initial decline after the peak rather than the whole tail, we get a best fit of $1+z=1.32$ and $F_{\rm obs}/F_{\rm model}=2.17\times 10^{-33}$. We show in Figure \ref{fig:lc} the separate fits to the entire lightcurve and the first 30 seconds, and we also include the model lightcurve fitted only by matching the peak fluxes, with the value for the redshift of $1+z=1.26$, as assumed by Heger et al.~(2007) . We see that while the redshift is sensitive to the details of the fitting, the normalization $F_{\rm obs}/F_{\rm model}$ is well-determined. Therefore, here we use the normalization, but leave fits to the shape of the entire lightcurve to future work when a greater number of simulations are available.

\begin{figure}
\includegraphics[width=\columnwidth]{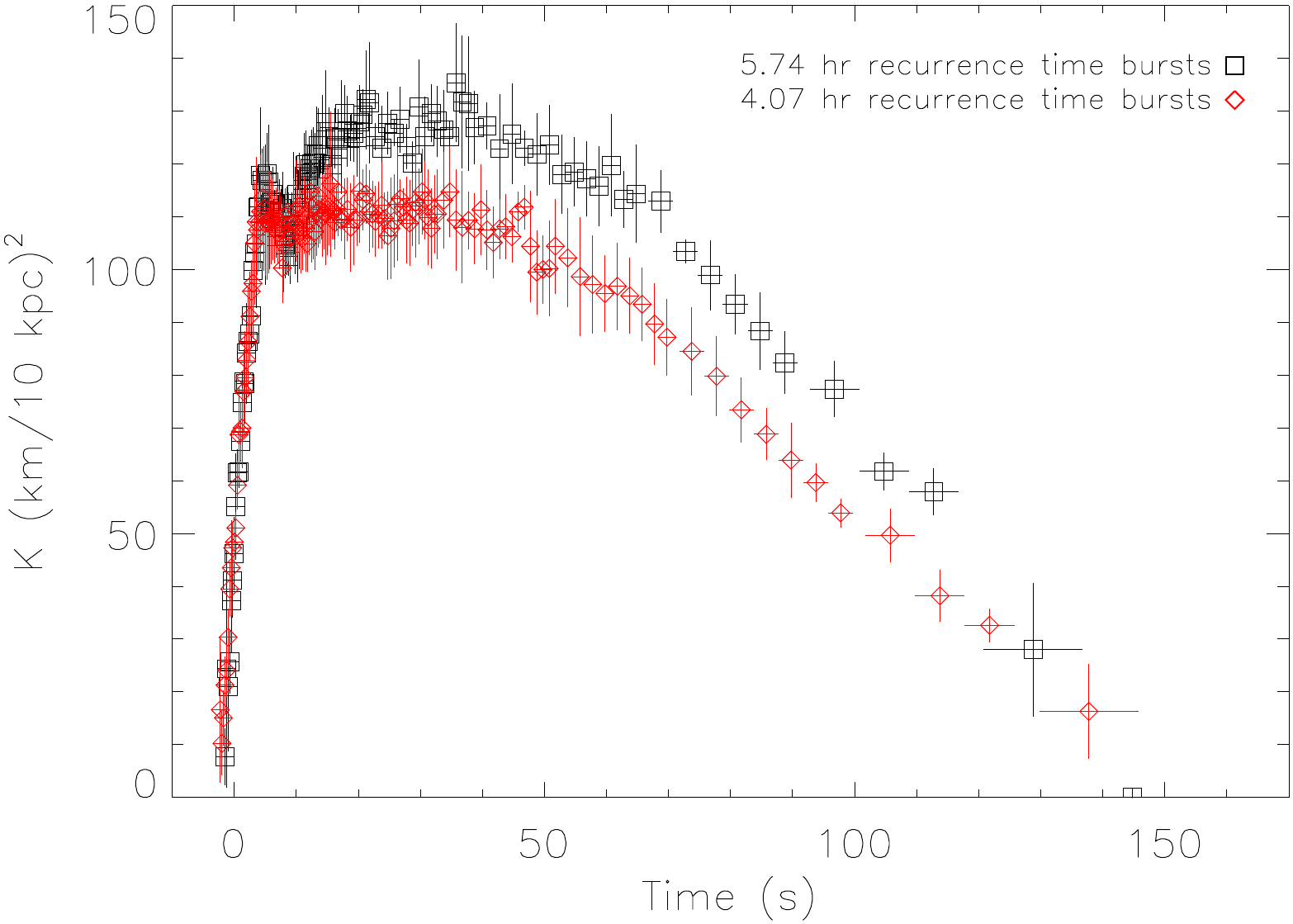}
\caption{Blackbody normalization for average burst profiles with $t_{\rm recur}=5.74$ (black squares) and $4.07\ {\rm hr}$ (red diamonds). \label{fig:Knorm}}
\end{figure}

A second constraint comes from spectral fitting. Fitting the observed burst spectrum with a blackbody gives the total flux $F_{\infty}$, color temperature $T_{c,\infty}$, and the  blackbody normalization
\begin{equation}\label{eq:K}
K=\left({F_\infty\over \sigma T_{c,\infty}^4}\right)={R^2_\infty\over d^2 f_c^4}\xi_b^{-1}.
\end{equation}
The color correction factor $f_c=T_c/T_{\rm eff}$ takes into account the hardening of the burst spectrum compared to a blackbody at the same effective temperature $T_{\rm eff}$. We plot $K$ as a function of time in Figure \ref{fig:Knorm} for two average burst profiles, for recurrence times $5.74$ and $4.07\ {\rm hr}$ respectively.  Following the burst rise, which lasts for approximately 5 seconds, the normalization levels off until $\approx 60$ seconds into the burst, when the normalization drops dramatically over 100 seconds to only about 25\% of its original value. We discuss the variation of $K$ in the tail, and the difference in $K$ for the two different recurrence times in the next section. For our purposes here we find the mean value of $K$ for the $t_{\rm recur}=4.07$ hr profile during the period following the peak where it is constant, giving $K=110\pm 2\ ({\rm km}/10\ {\rm kpc})^2$. If we take into account the deadtime correction near the burst peak ($\approx6\%$), the value we find is consistent with the more detailed analysis of blackbody normalization in these bursts carried out by Galloway \& Lampe (2011).

Dividing equations (\ref{eq:Fratio}) and (\ref{eq:K}), $R/d$ and $\xi_b$ drop out, giving
\begin{eqnarray}\label{eq:fczcorr}
{f_c\over 1+z}  &=&  K^{-1/4}  \left({F_{\rm obs}\over F_{\rm model}}\right)^{1/4}\\&=&1.17\ \left({K\over 110\ ({\rm km}/10 {\rm kpc})^2}\right)^{-1/4}\left({F_{\rm obs}/F_{\rm model}\over 2.2\times 10^{-33}}\right)^{1/4}.
\end{eqnarray}
 We show model calculations of $f_c$ from Suleimanov et al.~(2011b) in the next section which typically have $f_c\approx 1.4$--$1.5$ during the phase where $K$ is relatively constant. For $f_c=1.4,1.5$ we get $1+z=1.19,1.28$. %This is a lower redshift than we get from the ``stretch'' of the lightcurve, but it just relies on getting the peak luminosity correct and the uncertainties in measured $K$ or fluxes are small because of the 1/4 scaling.
Note that as well as being independent of $d$ and $\xi_b$, the value of $1+z$ determined in this way is not very sensitive to the values of $K$ and $F_{\rm model}/F_{\rm obs}$ (proportional to the 1/4 power of each). For example, introducing an uncertainty in $F_{\rm model}/F_{\rm obs}$ of $\pm10$\% gives a prefactor in equation (\ref{eq:z}) of $1.17\pm 0.03$ or $1+z=1.28\pm 0.03\ (f_c/1.5)$.

There is one more constraint which comes from the agreement between the measured and model recurrence times, which effectively measures the local mass accretion rate $\dot m$ onto the star. We define  $\dot m$ to be the rest mass accretion rate at the stellar surface. Then the accretion flux as observed at infinity is
\begin{equation}\label{eq:FX}
F_X=\dot m \left({R\over d}\right)^2 \xi_p^{-1} {c^2z\over (1+z)^2}.
\end{equation}
Dividing equations (\ref{eq:Fratio}) and (\ref{eq:FX}) gives the observed quantity
\begin{equation}
{F_X\over \dot m c^2}{F_{\rm model}\over F_{\rm obs}} = \left({\xi_b\over \xi_p}\right) z,
\end{equation}
a direct measure of redshift, independent of distance, but dependent on the anisotropy parameter ratio $\xi_p/\xi_b$. The accretion rate in model A3 of Heger et al.~(2007) was $\dot m=7980\ {\rm g\ cm^{-2}\ s^{-1}}$, and the measured persistent flux in 2000 was $F_X=2.91\pm0.03\times 10^{-9}\ {\rm erg\ cm^{-2}\ s^{-1}}$, giving 
\begin{equation}\label{eq:z}
z = 0.18\pm0.02\ \left({\xi_p\over \xi_b}\right).
\end{equation}
An alternative way to derive this result is to match the theoretical $\alpha$ value, the ratio of persistent fluence between bursts to burst fluence, to the observed value. The models of Heger et al.~(2007) with recurrence time of 4 hours have a theoretical value $\alpha_{\rm model} = \Delta M c^2z_{\rm model}/E_{\rm burst}(\xi_b/\xi_p)=55\ (\xi_b/\xi_p)$, where $E_{\rm nuc}$ is the burst energy, $\Delta M$ the ignition mass, and a redshift $z_{\rm model}=0.26$. The observed $\alpha$ at the same recurrence time is $\alpha_{\rm obs}\approx 37$ (Fig.~2 of Heger et al.~2007), giving $z=z_{\rm model}(\alpha_{\rm obs}/\alpha_{\rm model})=0.17 (\xi_p/\xi_b)$, in agreement with the value in equation (\ref{eq:z}).

If the anisotropy parameters were known, equations (\ref{eq:z}), (\ref{eq:Fratio}) and (\ref{eq:K}) uniquely determine the three quantities $1+z$, $f_c$ and $R/d$. However, as noted in \S 2, the anisotropy parameters are not well-constrained. In particular, equation ({\ref{eq:z}) for the redshift is not constraining once the large uncertainty in $\xi_p/\xi_b$ is taken into account. The main result of this section is therefore the relation between $1+z$ and $f_c$ in equation (\ref{eq:fczcorr}), since it is independent of $d$ and $\xi_b$.

In the next section, we constrain the Eddington flux by fitting the burst cooling tracks to spectral models. We can use the model lightcurve to say something about the expected value of $F_{\rm Edd}$, the observed flux which corresponds to the Eddington flux at the surface of the star. In the model the Eddington flux locally is $F_{\rm Edd}=cg/\kappa=0.882\times 10^{25} g_{14}$ erg cm$^{-2}$ s$^{-1}$ for $X=0.7$, giving $F_{\rm model,pk}/F_{\rm model,Edd}=1.46/g_{14}$. Scaling the observed peak flux, the Eddington flux as observed at infinity should be $1.95\times 10^{-8} g_{14}(1.7/1+X)$ \fluxcgs, or $F_{\rm Edd}/(10^{-8}$ \fluxcgs $)\,=1.95,3.89,7.79$ for $\log_{10} g=14.0,14.3,14.6$.

\begin{figure*}[htbp]
\includegraphics[width=\columnwidth]{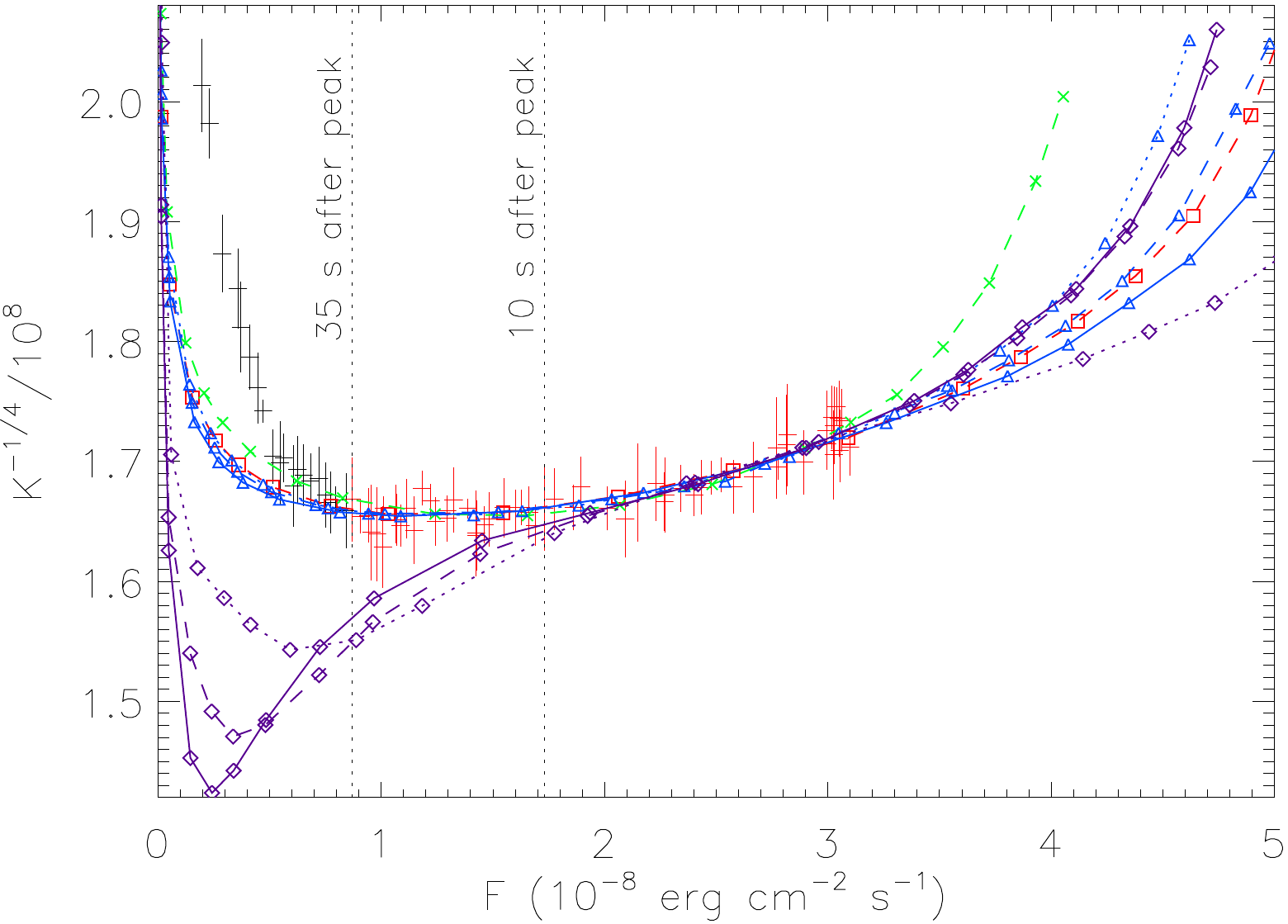}
\includegraphics[width=\columnwidth]{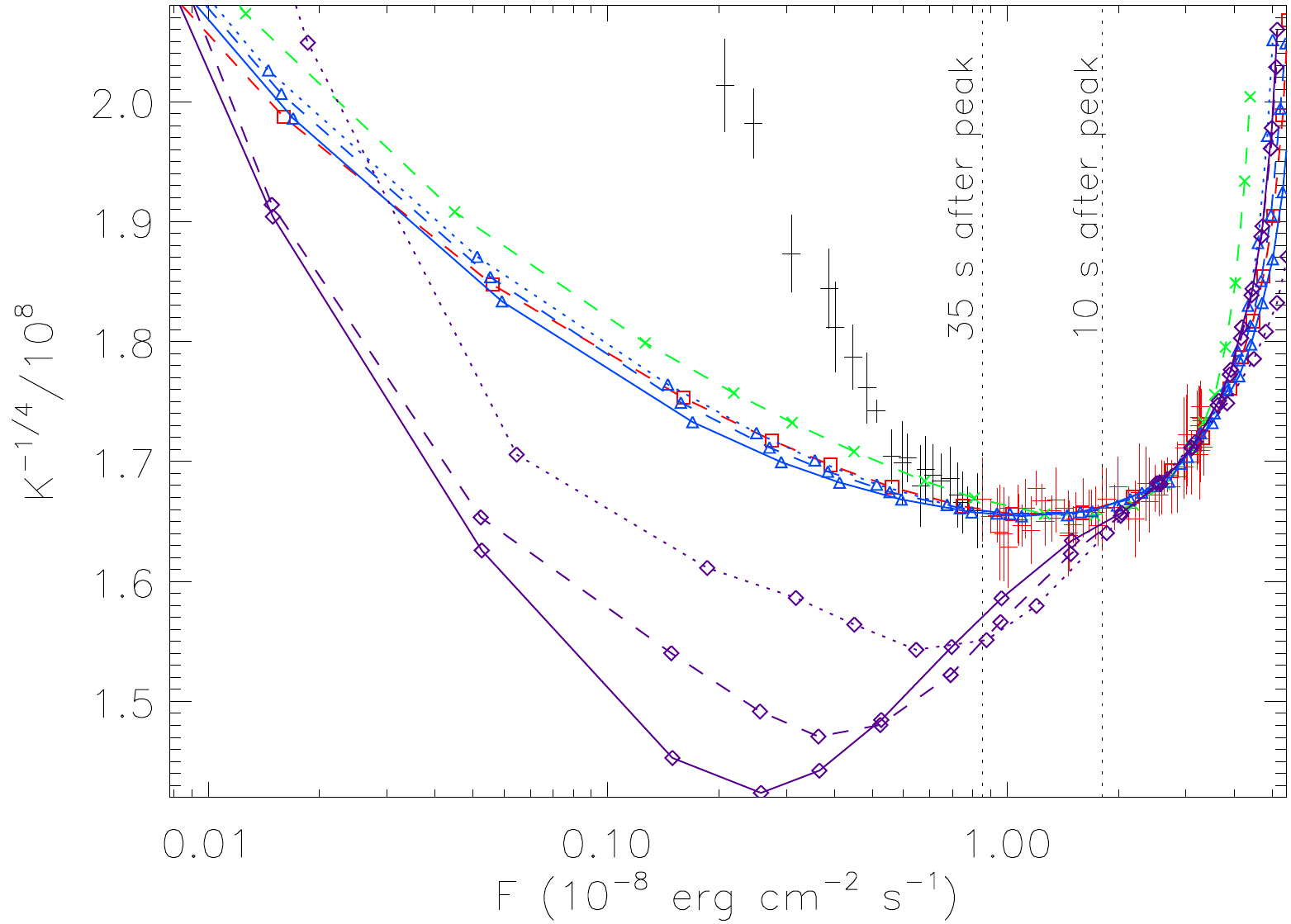}
\caption{Best fits to the theoretical $f_c-F/F_{\rm Edd}$ curves for a range of compositions. The crosses represent the all the data points from the burst peak onwards, and those in red representing only the first 35 seconds after the peak. Two vertical dotted lines represent the fluxes at $t=10,35\,$s after the burst peak. The compositions are solar H/He abundance with solar metallicity (diamonds connected by purple lines) or 1\% solar metallicity (triangles, blue lines), pure H (squares, red line) and pure He ($\times$ symbols, green line). Dotted, dashed and solid lines represent surface gravities of $\log_{10}(g)=14.0$, 14.3 and 14.6, respectively.\label{fig:fit}}
\end{figure*}

\section{Comparison with spectral models}

We now turn to fitting theoretical calculations of the color-correction factor $f_c$ to the data. First we describe the fitting procedure and results (\S 5.1) and then the constraints on neutron star parameters, in particular an upper limit on $R_\infty$ (\S 5.2). In \S 5.3 we discuss the variation of $K$ with accretion rate (Fig.~2) in the context of the spectral models.

\subsection{Fit for $A$ and $F_{\rm Edd}$}

Suleimanov et al.~(2011b) calculated $f_c$ as a function of $F/F_{\rm Edd}$ for a range of surface gravities and atmospheric compositions, and discussed how these models could be applied to data. We follow their analysis, and fit the theoretical $f_c$-$F/F_{\rm Edd}$ curves to the observed relation between $K^{-1/4}$ and flux $F$. The fitting parameters are $F_{\rm Edd}$ and $A=K^{-1/4}/f_c$. Comparing with equation (\ref{eq:K}), we see that 
\begin{equation}
A=\left({R_\infty\over d}\right)^{-1/2}\xi_b^{1/4}.
\end{equation}
If $f_c$ was a constant independent of flux, and $K$ was constant in the cooling tail of the burst, fitting for $A$ would be equivalent to using the measured normalization $K$ and a value of $f_c$ to extract $R_\infty/d$. Instead, here we are using the entire cooling track to obtain $A$. In addition, by fitting the shape of the cooling track we can obtain the overall flux scale $F_{\rm Edd}$ even though the burst itself does not reach Eddington luminosity. 

We start by fitting the data from \src\ with recurrence time $5.74$ hr to the different models from Suleimanov et al.~(2011b). Below we use the fits to obtain an upper limit on $R_\infty$, which motivates us to start with the bursts with the largest value of $K$ and therefore larger $R_\infty$ values. At the end of this section, we discuss whether it is possible to include the $4.07\ {\rm hr}$ recurrence time bursts which have smaller values of $K$ (Fig.~2) in a consistent picture. When fitting, for simplicity we calculate $\chi^2$ based on comparing $K^{-1/4}$ and $f_c$, and do not include the errors in the flux measurement. This seems reasonable because the observational error in the overall flux scale is $\approx \delta F/\sqrt{N}$, where the individual flux error $\delta F\approx 10^{-9}\ {\rm erg\ cm^{-2}\ s^{-1}}$, smaller than the overall uncertainty in the parameter $F_{\rm Edd}$ that we obtain from our fits.

Suleimanov et al.~(2011b) calculate spectral models for pure H and pure He atmospheres, and solar H/He fractions with different metallicity. Based on the lightcurve comparison and energetics, Galloway et al.~(2004) and Heger et al.~(2007) concluded that the accreted layer has solar metallicity and a substantial amount of hydrogen (a solar H/He ratio in their models). Here we fit to the full range of models from Suleimanov et al.~(2011b) to investigate how changing composition affects our derived limits on neutron star parameters. Also, the photospheric abundances could be different from the abundances near the base of the layer where the X-ray burst ignites. For example, the metallicity in the burning layer could be enhanced by partially burned fuel left over from a previous burst. The hydrogen fraction in the accreted material could be lower than solar, and so it is useful to consider the pure He limit as a limiting case when the hydrogen fraction at the photosphere is reduced.

Figure \ref{fig:fit} shows example fits to the 5.74 hr recurrence time bursts. By varying $A$ and $F_{\rm Edd}$, we are able to obtain good fits for fluxes down to approximately 1/3 of the peak flux. At lower fluxes, the behavior of the model and observations is qualitatively similar, in that $f_c$ rises rapidly at low fluxes, but the detailed behavior does not match the models. At late times or low fluxes, $K^{-1/4}$ rises more rapidly than predicted. We therefore confine the fit to the initial part of the cooling tail and use it to derive $A$ and $F_{\rm Edd}$. To do this, we fit $\Delta t_{\rm fit}$ seconds of data starting at the time of peak flux. The time $\Delta t_{\rm fit}$ is chosen so that as much data is included in the fit as possible while still giving an adequate fit (with the late time data excluded, we find reduced $\chi^2$ values in the range 0.23--0.47 for the models listed in Table 1). For all except the solar metallicity models we take $\Delta t_{\rm fit}=35\ {\rm s}$, corresponding to fluxes down to $\approx 1/3$ of the peak flux. The solar metallicity models begin to deviate from the data after $\Delta t_{\rm fit}=10\ {\rm s}$ (about 1/2 of the peak flux) because of the dip in $f_c$ at low fluxes. 

The results of the fits for different spectral models are listed in Table \ref{tab:fcfits}. We used Markov Chain Monte Carlo implemented with the Metropolis-Hastings algorithm (Gregory~2005, Chap.~12) to sample the parameter space and find the distributions for the fitting parameters, $A$ and $F_{\rm Edd}$. Those distributions were then each fitted by a Gaussian profile in order to derive their respective central values and $1\sigma$ uncertainties. In certain cases, as noted in Table \ref{tab:fcfits}, the distributions were not well described by a single Gaussian profile, due to the presence of more than one peak. In those cases, we fit a Gaussian profile to the peak at the lowest values of $F_{\rm Edd}$ and $A$. Note that these parameters are correlated since an increase (decrease) in $A$, which moves the model curves upwards (downwards) with respect to the observations, can be offset by a corresponding increase (decrease) in $F_{\rm Edd}$ which moves the curves rightwards (leftwards). The range of $F_{\rm Edd}$ for all the fitted models is from $4.1$ to $7.4\times 10^{-8}\ {\rm erg\ cm^{-2}\ s^{-1}}$, which lies within the range of $F_{\rm Edd}$ from the Heger et al.~(2007) models used in section \S 4. Excluding the 0.1$Z_{\Sun}$, $\log_{10}g=14.0$ model, the range of $F_{\rm Edd}$ is relatively narrow; $4.1$ to $5.9\times 10^{-8}$ \fluxcgs.

The behavior at low fluxes is shown in more detail in the second panel of Figure \ref{fig:fit} which has a logarithmic flux axis. The slope of the increase in $f_c$ with decreasing flux is steeper in the data than in the models for low metallicity models. For solar metallicity models, the slopes are similar. This would enable a good fit of the whole data set to those models, particularly at low fluxes, but only if $F_{\rm Edd}$ is in the range $15$--$25\times 10^{-8}$ \fluxcgs. In fact, even with the restriction of using only 10 seconds of data following the burst peak we found other adequate fits, as separate local $\chi^2$ minima, in that range of $F_{\rm Edd}$. This is much larger than expected and as can be seen from the relations derived below, would give very small limits on $R_\infty$, and so we do not consider these fits further.

\begin{deluxetable*}{cccccccccc}
\tablewidth{0pt}
\tablecaption{Fits of spectral models to the burst cooling tail }
\tablehead{\colhead{Composition} & \colhead{$\log_{10} g$} & \colhead{$\Delta t_{\rm fit}$\tablenotemark{a}} & \colhead{$A$} & \colhead{$F_{\rm Edd}$}&  $\chi^2_{\rm reduced}(d.o.f)$ & \multicolumn{4}{c}{Upper limits}\\
\colhead{} & \colhead{} & \colhead{(s)} & \colhead{$(10^8)$} & \colhead{($10^{-8}\ {\rm erg\ cm^{-2}\ s^{-1}}$)} &  
&\colhead{$d\xi_b^{1/2}$ (kpc)} &\colhead{$R_\infty$ (km)} & \colhead{$M_{\rm max} (M_\odot)$\tablenotemark{b}} &\colhead{$R_{M>M_\odot}$ (km)\tablenotemark{c}} \label{tab:fcfits}}
\startdata
Pure H & 14.3 & 35 & 1.144$\pm 0.002$ & 5.15$\pm0.20$ & 0.27 (59)  & 4.3 & 10.2 & 1.3 & 8.2\\ 
Pure He & 14.3 & 35 & 1.186$\pm 0.002$ & 4.14$\pm0.09$ & 0.26 (59) & 9.7 & 21.5 & 2.8 & 19.8\\
0.01$Z_{\Sun}$\tablenotemark{d} & 14.0 & 35 & 1.138$\pm0.002$ & 4.71$\pm0.14$ & 0.24 (59) & 5.5 & 13.2 & 1.7 & 11.3\\
0.01$Z_{\Sun}$\tablenotemark{d} & 14.3 & 35 & 1.154$\pm 0.002$ & 5.08$\pm0.15$ & 0.23 (59) & 5.0 & 11.6 & 1.5 & 9.7\\
0.01$Z_{\Sun}$\tablenotemark{d} & 14.6 & 35 & 1.167$\pm 0.002$ & 5.43$\pm0.19$ & 0.26 (59) & 4.6 & 10.5 & 1.4 & 8.4\\
0.1$Z_{\Sun}$\tablenotemark{d} & 14.0 & 35 & 1.170$\pm 0.003$ & 7.35$\pm0.66$ & 0.47 (59) & 4.0 & 9.0 & 1.2 & 6.8\\
0.1$Z_{\Sun}$\tablenotemark{d} & 14.3 & 35 & 1.168$\pm 0.004$ & 5.80$\pm0.41$ & 0.45 (59) & 4.5 & 10.2 & 1.3 & 8.1\\
0.1$Z_{\Sun}$\tablenotemark{d} & 14.6 & 35 & 1.174$\pm 0.003$ & 5.75$\pm0.31$ & 0.36 (59) & 4.4 & 9.9 & 1.3 & 7.8\\
$Z_{\Sun}$\tablenotemark{d,}\tablenotemark{e} & 14.0 & 10 & 1.178$\pm 0.020$ & 5.92$\pm1.22$ & 0.36 (31) & 5.3 & 12.4 & 1.6 & 10.6\\
$Z_{\Sun}$\tablenotemark{d,}\tablenotemark{e} & 14.3 & 10 & 1.159$\pm0.017$ & 4.81$\pm1.08$ & 0.29 (31) & 5.6 & 13.2 & 1.7 & 11.3\\
$Z_{\Sun}$\tablenotemark{d,}\tablenotemark{e} & 14.6 & 10 & 1.164$\pm 0.011$ & 4.84$\pm0.52$ & 0.25 (31) & 5.4 & 12.6 & 1.6 & 10.7
\enddata
\tablenotetext{a}{We fit to data from the time of peak luminosity until $\Delta t_{\rm fit}$ seconds later.}
\tablenotetext{b}{The maximum neutron star mass consistent with the upper limit on $R_\infty=R(1+z)$.}
\tablenotetext{c}{The upper limit on radius assuming a lower limit on mass $M>1\ M_\odot$.}
\tablenotetext{d}{The composition is solar H/He abundance plus the indicated proportion of solar metallicity.}
\tablenotetext{e}{These fits yielded more than one local $\chi^2$ minima. Here, we report only the minima located at the lowest values of $F_{\rm Edd}$ and $A$. See text for more details.}
%\tablenotetext{f}{The lowest $\chi^2$ fit parameters $A$ and $F_{\rm Edd}$ did not correspond to the peak of their respective fit distributions. Here we quote the best fit $A$ but give the range implied from the value of $\sigma$ from the fitted Gaussian.}
\end{deluxetable*}

\subsection{Upper limit on distance and $R_\infty$}

A measurement of $A$ and $F_{\rm Edd}$ translates into typically two values for $M$ and $R$ as follows. 
When the flux at the surface of the star is at the local Eddington flux $cg/\kappa$, the observed flux is
\begin{equation}
F_{\rm Edd}={GMc\over \kappa d^2}{1\over 1+z}\xi_b^{-1},
\end{equation}
where we take the opacity to be $\kappa=0.2\ {\rm cm^2\ g^{-1}}\ (1+X)$, as used in Suleimanov et al.~(2011b).
Following Steiner et al.~(2010) we define the quantities
\begin{equation}\label{eq:alphadef}
\alpha \equiv {\kappa d\over c^3}F_{\rm Edd}A^2 \xi_b^{1/2} = {u\over 2}(1-u)
\end{equation}
\begin{equation}
\gamma \equiv {c^3\over \kappa}{1\over A^4F_{\rm Edd}} = {R\over (u/2)(1-u)^{3/2}} 
\label{eq:gamma}
\end{equation}
where $u=2GM/Rc^2$. These definitions differ slightly from those of Steiner et al.~(2010) in that they include the anisotropy parameter $\xi_b$. Then
\begin{equation}
u = {1\over 2}\pm {1\over 2}\left(1-8\alpha\right)^{1/2}
\end{equation}
\begin{equation}
R=\alpha\gamma\left(1-u\right)^{1/2}.
\label{eq:R}
\end{equation}
To calculate $\alpha$ and $\gamma$ from $A$ and $F_{\rm Edd}$ obtained from the fits, we require distance $d$ and composition $X$. A given $\alpha$ and $\gamma$ then give two solutions for $M$ and $R$. 

We treat the $A$ and $F_{\rm Edd}$ values as given independently of the derived $M$ and $R$. In fact the derived $A$ and $F_{\rm Edd}$ values depend on the gravity assumed for the spectral models. 
The color correction $f_c$ decreases with increasing gravity in the models of Suleimanov et al. (2011b) (see their Fig.~5). This implies an increasing $A$ with gravity, which we find in our results for the $Z=0.01Z_{\Sun}$ models. A similar trend is not seen in the $Z=0.1Z_{\Sun}$, $Z_{\Sun}$ models. The color correction still decreases with gravity, but the shapes of the models change in such a way as to favor smaller values of $F_{\rm Edd}$, which, given the correlation between the two fitting parameters, leads to values for $A$ that are smaller than expected for the higher gravities. The value of $F_{\rm Edd}$ also increases with gravity for the $Z=0.01Z_{\Sun}$ models, because the slope of $f_c$ with $F/F_{\rm Edd}$ steepens with increasing gravity, requiring a larger value of $F_{\rm Edd}$ to agree with observed $K^{-1/4}$--$F$ slope. 

Overall, we found that the fits are sensitive to the detailed shape of the atmosphere models. This could be due to some small irregularities in the model slopes attributable to the coarseness of the flux grid on which the color corrections are evaluated, and the limited range of fluxes spanned by the bursts we are analyzing.

Table \ref{tab:fcfits} shows that for a given metallicity, varying the surface gravity changes changing gravity from $\log_{10} g=14.0$ to $14.6$ changes $A$ by up to $\approx 3$\% for low metallicity, and $F_{\rm Edd}$ by up to $\approx 20$\%. The resulting changes in the limit on $R_\infty$ are $\approx 30$\% for low metallicity models. This gives a measure of the error introduced by not carrying out a self-consistent fit in which the gravity of the spectral model used and derived $M$ and $R$ are consistent.

Equation (\ref{eq:alphadef}) shows that real-valued solutions for $u$ require $\alpha\leq 1/8$ (Steiner et al.~2010). As emphasized by Suleimanov et al.~(2011a, 2011b), this gives an upper limit on the distance for which solutions are possible,
\begin{eqnarray}\label{eq:dlim}
\xi_b^{1/2} d &\leq&  {1\over 8} {c^3\over \kappa}{1\over F_{\rm Edd}A^2}
\\&=&  5.6\ {\rm kpc}\ \left({A_8\over 1.2}\right)^{-2}\left({F_{{\rm Edd},-8}\over 4.0}\right)^{-1}\left({1+X\over 1.7}\right)^{-1},
\end{eqnarray}
where $A_8=A/10^{8}$ and $F_{{\rm Edd},-8}=F_{\rm Edd}/(10^{-8}$ \fluxcgs). It also provides a limit on $R_{\infty}=R(1+z)$. To see this, note that the neutron star radius is
\begin{eqnarray}
R_\infty&=&R(1+z)=\alpha\gamma\\
&=&12.0\ {\rm km}\ \alpha\ \left({A_8\over 1.2}\right)^{-4}\left({F_{{\rm Edd},-8}\over 4.0}\right)^{-1}\left({1+X\over 1.7}\right)^{-1}.\label{eq:rinflim}
\end{eqnarray}
where the definition of $\gamma$ from equation (\ref{eq:gamma}) was used and $(1-u)^{1/2}$ was substituted with $(1+z)^{-1}$. An upper limit on $R_\infty$ is obtained by setting $\alpha=1/8$ in equation (\ref{eq:rinflim}).

The upper limits on $\xi_b^{1/2} d$ and $R_\infty$ are given in Table \ref{tab:fcfits}. To calculate them we use equations (\ref{eq:dlim}) and (\ref{eq:rinflim}) with 95\% lower limits on the quantities $A^2F_{\rm Edd}$ and $A^4F_{\rm Edd}$ derived from our fits. A slightly different procedure is used for the cases where the fits yielded multiple $\chi^2$ minima. To derive the most conservative upper limits on $\xi_b^{1/2} d$ and $R_\infty$, we consider only the $\chi^2$ local minimum located at the lowest value of $F_{\rm Edd}$ and $A$, manifested as a distinct, Gaussian-like peak in the respective distributions for the quantities $A^2F_{\rm Edd}$ and $A^4F_{\rm Edd}$. Considering only the part of the Gaussian-like distribution lying below the peak value, we find the 90\% lower limits for $A^2F_{\rm Edd}$ and $A^4F_{\rm Edd}$. This is equivalent to taking the 95\% lower limit of the whole peak, but has the advantage of allowing us to isolate the $\chi^2$ minimum of interest from the rest of the distribution. As a check, we applied this procedure to model fits showing a single $\chi^2$ minimum, and found very small differences ($<1\%$) in the derived upper limits when compared to those found by considering the entire distributions. 

 An upper limit on $R_\infty$ implies an upper limit on the neutron star mass $M_{\rm  max}=c^2R_\infty/3^{3/2}G$ (at that mass the radius is $R_\infty/\sqrt{3}$), also given in Table \ref{tab:fcfits}. Note that the upper limits on $R_\infty$ and $d$ are correlated. Since $\xi_b^{1/2}d_{\rm lim}=\gamma A^2/8$ (compare eqs.~[\ref{eq:gamma}] and [\ref{eq:dlim}]), we can rewrite equation (\ref{eq:rinflim}) as
\begin{equation}
R_\infty<12.0\ {\rm km}\ \left({\xi_b^{1/2} d_{\rm lim}\over 5.6\ {\rm kpc}}\right)\left({A_8\over 1.2}\right)^{-2},
\end{equation}
a larger distance limit allows larger radii.

For solar abundance of hydrogen at the photosphere, we find $\xi^{1/2}d\lesssim 4.0$--$5.6\ {\rm km}$ and $R_\infty<9.0$--$13.2$ km. This represents quite stringent limits on the neutron star mass and radius. For this range of $R_\infty$, the maximum neutron star mass is in the range $1.2$ to $1.7\ M_\odot$. If we consider a lower mass limit of $1\ M_\odot$, the neutron star radius must be smaller than $R(1\ M_\odot)=6.8$--$11.3\ {\rm km}$ (the individual values for each model are given in Table \ref{tab:fcfits}).

\begin{figure}
\includegraphics[width=\columnwidth]{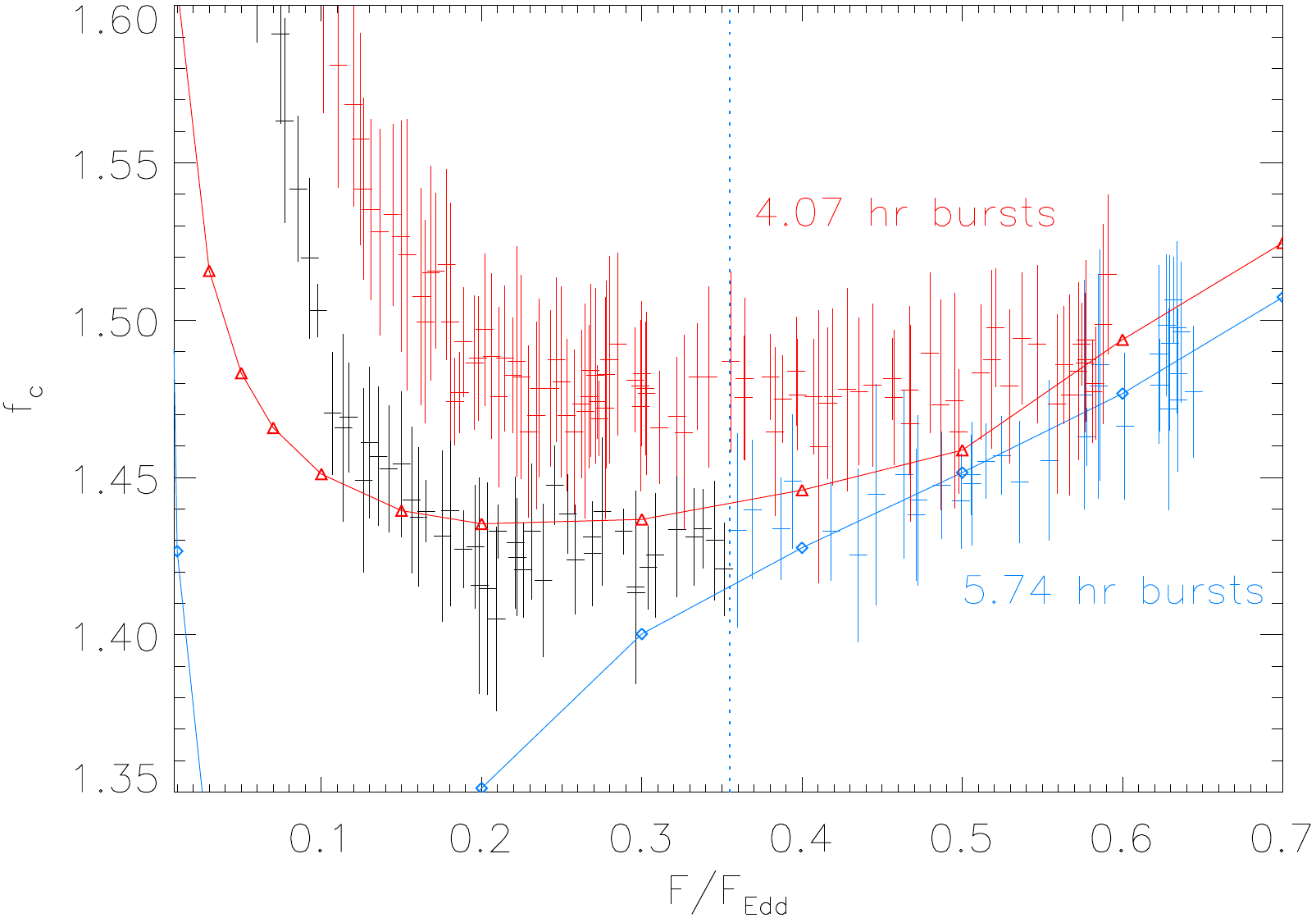}
\includegraphics[width=\columnwidth]{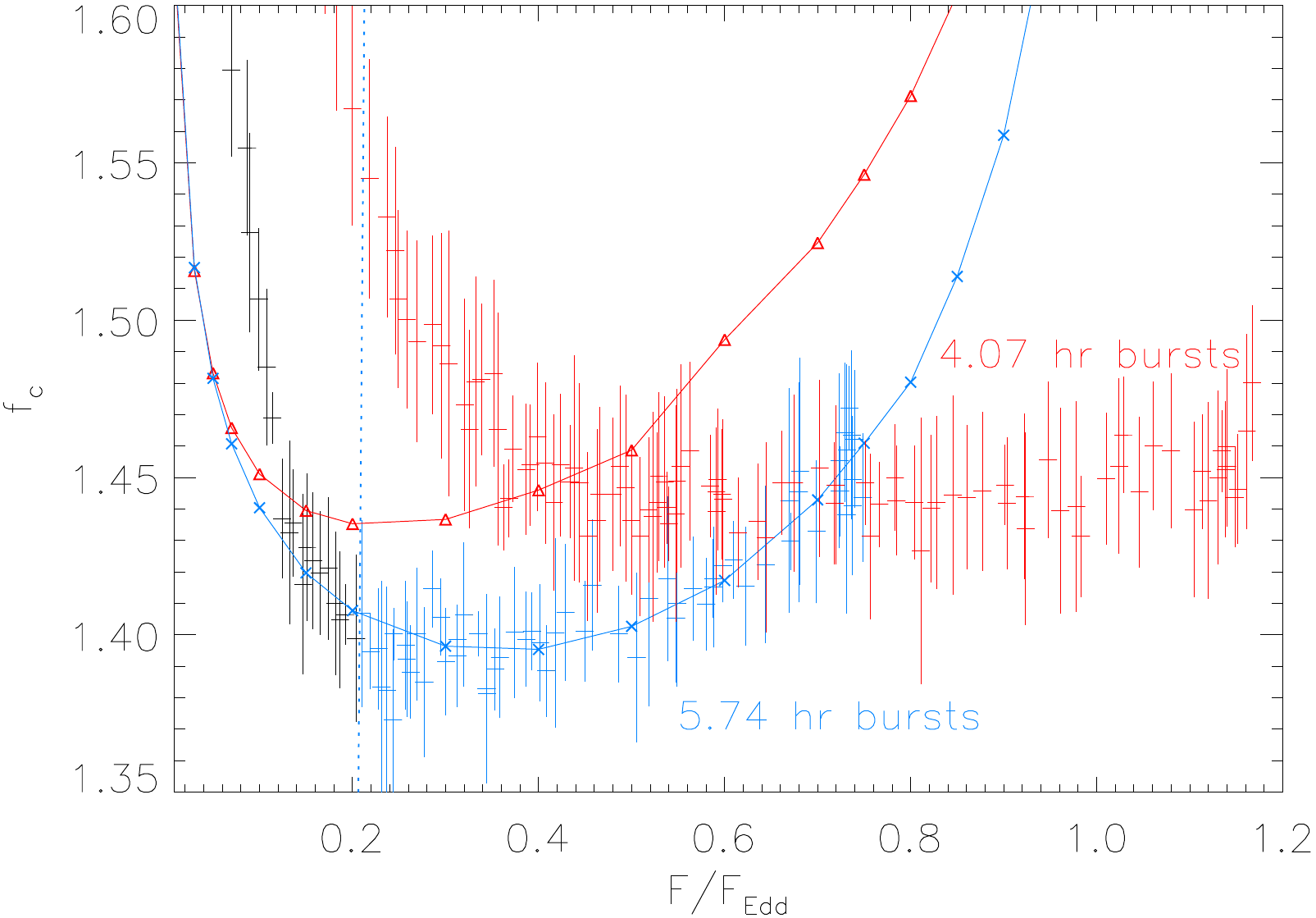}
\caption{Top panel: The solar metallicity fit ($\log_{10}g=14.3$, linked blue diamonds) that reproduces the first part of the cooling track for the $t_{\rm recur}=5.74$ hr bursts (blue crosses, and delimited from the data exluded from the fit by a dotted vertical blue line) is plotted together with the data for the $t_{\rm recur}=4.07$ hr bursts (red crosses) and the low metallicity $Z=0.01Z_\odot$ model (linked red triangles) at the same $A$ and $F_{\rm Edd}$ as the solar metallicity model. Bottom panel: A pure He spectral model (linked blue $\times$ symbols) fit to the 5.74 hr bursts (blue crosses) and a low metallicity solar H/He composition model (linked red triangles) at the same $A$ and $F_{\rm Edd}$ adjusted for the different hydrogen fraction. In both panels, the respective symbols show the points where the atmospheric models were calculated.\label{fig:varycomposition}}
\end{figure}

\subsection{Variation of $K$ with accretion rate}

Figure \ref{fig:Knorm} shows that the bursts with recurrence times of $4.07\ {\rm hr}$ have significantly smaller values of $K$ than the 5.74 hr bursts, by $\approx 20$\% (see Galloway \& Lampe 2011 for a detailed discussion of the variation of $K$ in the sample of bursts from \src). Variations in $K$ between bursts has been seen in other sources. For example, Damen et al.~(1989) found that the blackbody temperature (evaluated at a fixed flux level) depended on burst duration. They suggested that variations in chemical composition at the photosphere and the resulting changes in color correction might explain the changing blackbody temperature (and therefore normalization). 

We investigate two possible composition variations: changing metallicity with solar H/He abundance, and changing the hydrogen fraction. First, we consider solar H/He abundance and changing metallicity. Suleimanov et al.~(2011b) show that $f_c$ drops with increasing metallicity. Therefore we fit the solar metallicity model to the $5.74\ {\rm hr}$ bursts to determine values of $A$ and $F_{\rm Edd}$ (as given in Table 1). These values are then used to compare a low metallicity model to the $4.07\ {\rm hr}$ data. This comparison is shown in the top panel of Figure \ref{fig:varycomposition}. The low metallicity model lies below the $4.07$ hr data, showing that the difference in $K$ cannot be explained by a decrease in metallicity from solar to a fraction of solar.

Second, we consider a change in hydrogen fraction at the photosphere. The lower panel of Figure \ref{fig:varycomposition} shows the pure He atmosphere fit for the 5.74 hour bursts (see Table 1), and a low metallicity solar H/He abundance model for the 4.07 hr bursts in which we use the same value of $A$ determined by the pure He atmosphere fit, but decrease the derived $F_{\rm Edd}$ by a factor of $1+X=1.7$ to account for the difference in Eddington flux with composition. This plot shows that the change in $f_c$ in going from pure He to solar H composition is enough to account for the variation in $K$ observed. However, the solar composition model does not match the 4.07 hr data in terms of location on the $F/F_{\rm Edd}$ axis. Another way to say it is that if we fit the 4.07 hr data with a solar composition model, the required $F_{\rm Edd}$ would be larger than for the $5.74$ hr data, instead of being a factor $1+X$ times smaller, as is required for simultaneous fits. Furthermore, we see in the lower panel of Figure \ref{fig:varycomposition} that reducing the derived $F_{\rm Edd}$ by a factor of $1+X=1.7$ for the 4.07 hr bursts implies that the peak flux for those bursts exceeds the Eddington limit, which is known not be the case. Therefore a consistent explanation of the variation in $K$ in terms of changing H fraction at the photosphere is not possible.

\begin{figure}
\plotone{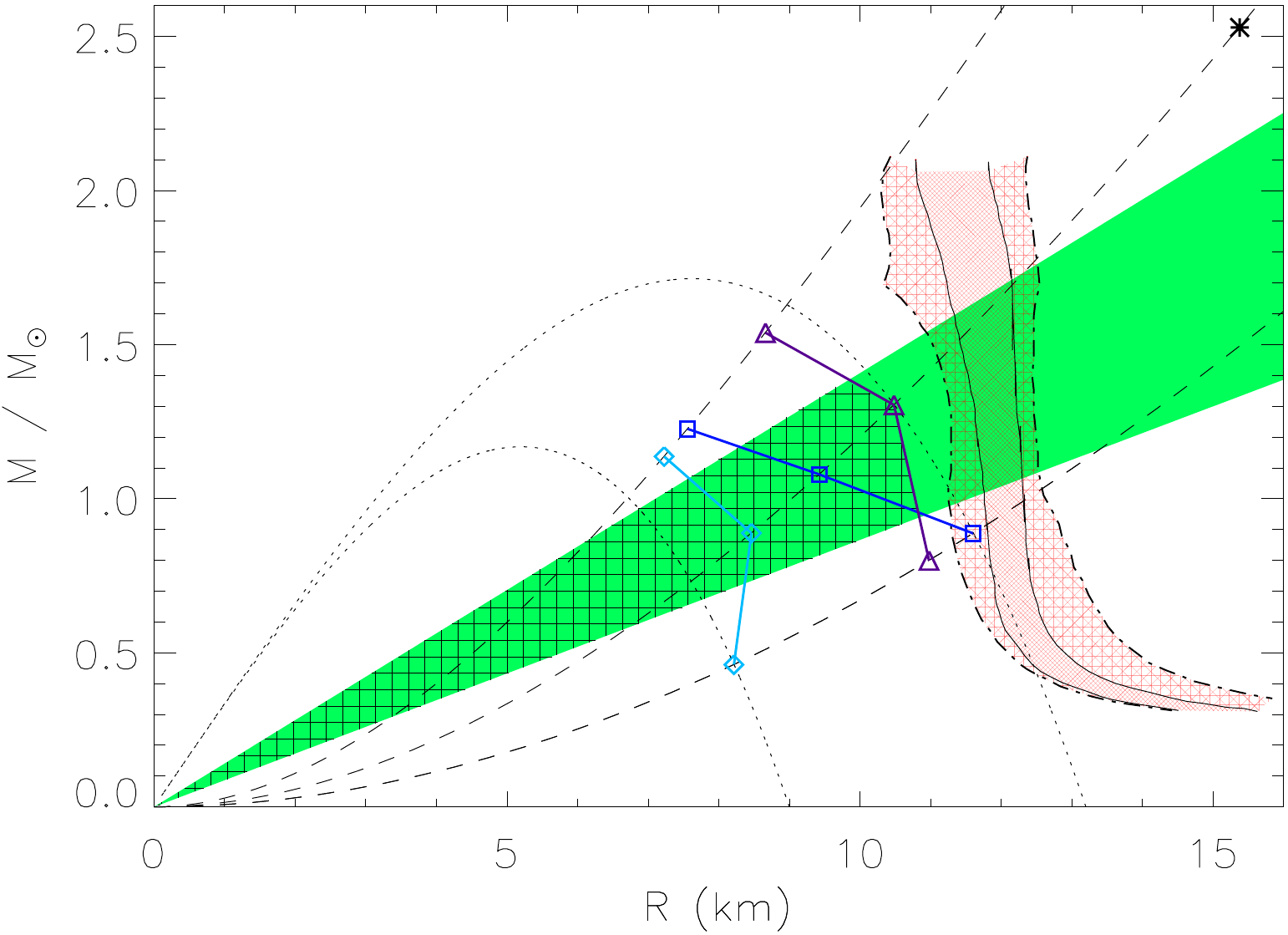}
\caption{Summary of distance-independent constraints in the neutron star mass-radius plane. The dashed curves are lines of constant surface gravity $\log_{10}(g)=14.0,14.3,14.6$ (bottom to top), values at which the spectral models were evaluated. In green, we show the redshift from eq.~(\ref{eq:fczcorr}) for $f_c=1.4$--$1.5$ and an assumed 10\% uncertainty in $F_{\rm obs}/F_{\rm model}$. The squares (dark blue), diamonds (light blue) and triangles (purple) represent the upper limits on $R_{\infty}$ computed from fits to the solar H/He abundance models with $0.01Z_{\sun}$, $0.1Z_{\sun}$ and $Z_{\sun}$ metallicities, respectively, each at a specific surface gravity. The upper limit on $R_{\infty}$ for the pure Helium atmosphere model ($\log=14.3$) is also shown as a black asterisk. Two constant $R_{\infty}$ curves are plotted as dotted lines for the highest and lowest values found within solar H/He abundance models. The region hashed in black represents what is allowed by the combination of the constraints derived from the fit to the burst lightcurve and spectral fits to solar H/He abundance models. These constraints are independent of the source distance and anisotropy parameters $\xi_b$, $\xi_p$. The region in red represents the mass-radius relation derived by Steiner et al.~(2010) (based on the $r_{\rm ph}\gg R$ assumption), with the 1$\sigma$ and 2$\sigma$ regions delimited by solid and dot-dashed lines, respectively. \label{fig:mr}}
\end{figure}

\section{Summary and Discussion}

We have compared lightcurve and spectral models with observations of Type I X-ray bursts from \src. Here we summarize the main conclusions and discuss our results further.

A general point is that anisotropy in the burst emission enters as an additional uncertainty in any derived quantity that depends on distance. Since it changes the relation between the source luminosity and observed flux, the anisotropy parameter $\xi_b$ (defined in \S 3) always enters in combination with distance as $\xi_b^{1/2} d$. Even in cases where distance to a source can be accurately determined, the anisotropy introduces an effective uncertainty of up to a factor of 20-30\%. Anisotropy could be a smaller effect for PRE bursts if the inner disk is disrupted during the burst and intercepts a smaller amount of radiation than a disk extending all the way to the stellar surface. Nonetheless, it remains a source of systematic error on derived neutron star radii that needs to be investigated further. For \src, the limit $i<70^\circ$ from Homer et al.~(1998) gives $\xi_b^{-1/2}=0.9$--$1.2$. Given this uncertainty and the fact that the distance to \src\ is not well constrained, we focused on deriving limits on $M$ and $R$ that are independent of distance and anisotropy. 

The first of these constraints comes from using the model lightcurve from Heger et al.~(2007) to fix the overall luminosity scale of the observed bursts. We showed that this leads to a distance and anisotropy independent relation between the redshift $1+z$ and color correction factor $f_c$ (eq.~[\ref{eq:fczcorr}]) that depends weakly on the measured normalization $K$ and the ratio of observed and model peak fluxes. For a color correction between 1.4 and 1.5, which spans the range of values in Fig.~2 of Suleimanov et al.~(2011) for example, the inferred redshift is between $z=0.19$ and $0.28$. 

The second constraint comes from comparing the spectral evolution during the cooling tail with the spectral models of Suleimanov et al.~(2011b), which determines the Eddington flux $F_{\rm Edd}$ and the quantity $A=K^{-1/4}/f_c$. As noted by Suleimanov et al.~(2011b), for a given set of measured $F_{\rm Edd}$, $A$ parameters, there is an upper limit to the distance of the source beyond which there is no solution for $M$ and $R$. We point out here that measuring $A$ and $F_{\rm Edd}$ also places an upper limit on $R_\infty=R(1+z)$ (and therefore upper limits on $M$ and $R$ for a given source). This limit is independent of distance and anisotropy and depends only on the measured values of $A$ and $F_{\rm Edd}$ and the surface hydrogen fraction. For \src, atmospheric models with solar hydrogen fractions give $R_\infty<9.0$--$13.2$ km (Table \ref{tab:fcfits}) which implies a neutron star mass $M<1.2$--$1.7\ M_\odot$ and $R< 6.8$--$11.3\ {\rm km}$ assuming a lower mass limit of $1\ M_\odot$. The corresponding distance limits are $d<4.0$--$5.6\ {\rm kpc}\ \xi_b^{-1/2}$. 

Uncertainties associated with absolute flux calibration do not affect our results; they are equivalent to an incorrect measurement of the distance to the source, which our constraints are independent of.

The constraints on $M$ and $R$ are summarized in Figure \ref{fig:mr}. We show the upper limits on $R_\infty$ from Table \ref{tab:fcfits} for all the solar hydrogen composition models each plotted at the respective surface gravity and the pure Helium model with $\log g=14.3$, and the redshift range $1+z=1.16$--$1.31$ from equation (\ref{eq:fczcorr}) with $f_c=1.4$--$1.5$ and a 10\% uncertainty in the ratio $F_{\rm obs}/F_{\rm model}$. The limits on radii for the solar hydrogen composition are comparable to but a little lower than current theoretical expectations based on dense matter calculations which have radii of 10--13 km for neutron star equations of state that reach a maximum mass $>2 M_\odot$ (Hebeler et al.~2010; Gandolfi, Carlson, \& Reddy 2011).  The mass-radius relation found in Steiner et al.~(2010), derived from a set of photospheric radius expansion X-ray bursts and hydrogen atmosphere fits for transiently accreting neutron stars in quiescence, also lies at slightly larger radii than our $R_\infty$ limits for solar composition. It should be noted that Suleimanov et al.~(2011a) call into question the results of Steiner et al.~(2010) by suggesting that ``short'' PRE bursts should be excluded from analysis as they show smaller blackbody normalizations in the burst tail and also do not follow the theoretically expected spectral evolution. The implication is that the mass-radius relation derived in Steiner et al.~(2010) would shift to higher radii as a result of using the more reliable ``long'' PRE bursts, and thus farther away from our derived upper limits.

A smaller hydrogen fraction at the photosphere in \src\ would increase the $R_\infty$ limits and make them consistent with the theoretical calculations and the mass-radius curve from Steiner et al.~(2010). We can get an impression of what the upper limit on $R_\infty$ would be for an atmosphere with a reduced hydrogen fraction by first looking at the extreme case of the pure helium atmosphere, and its derived upper limit of $21.5$ km for $\log_{10}(g)=14.3$ (see Figure \ref{fig:mr}). Such an upper limit is consistent with theoretical calculations and the results from Steiner et al.~(2010). We can go one step farther and estimate the hydrogen fraction we would require to have such a consistency with previous results using Equation \ref{eq:rinflim}. Assuming a surface gravity of $\log_{10}(g)=14.3$, an upper limit on $R_\infty$ of $\sim$16 km or more would be required. Using values for $F_{\rm Edd}$ and $A$ averaged across the solar H/He model fits with $\log_{10}(g)=14.3$, we estimate that a hydrogen fraction of $X\approx0.5$ or less would be needed. We are assuming that such a spectral model would not differ too greatly in shape from the solar H/He models. Galloway et al.~(2004) find that the theoretical variations in burst properties with persistent flux between ignition models with X=0.7 and X=0.5 are largely indistinguishable. However, more burst lightcurve simulations would be necessary to establish whether agreement with observed lightcurves is still possible with a reduced accreted hydrogen fraction.

Given that the upper limit on $R_{\infty}$ depends on the color correction as $f_c^4$ (via $A$ in equation \ref{eq:rinflim}), even a small 5\% increase in the value of $f_c$ would yield a $\sim22\%$ increase in  the upper limit on $R_{\infty}$. Furthermore, Suleimanov et al.~(2011b) discuss large color correction factors of $f_c=1.6$--$1.8$ (cf. also Suleimanov \& Poutanen 2006) possibly arising from a spreading layer associated with accretion onto the neutron star equator.  Perhaps in \src\ something similar is happening, although the increase in color correction required is not as large. It is worth noting that changing the visible area, for example by blocking one hemisphere of the neutron star with the accretion disk during the burst, does not change the inferred limits on radius because the limit on $R_\infty$ is independent of the anisotropy factor $\xi_b$ (isotropic emission from only half the area is equivalent to setting $\xi_b=2$).

There are several points to keep in mind when looking at our derived constraints on $M$ and $R$. First, the constraints are only partly self-consistent in the sense that the lightcurve model used to fit the data does not have the same gravity as the  derived $M$ and $R$. Heger et al.~(2007) (and Woosley et al.~2004) used a specific choice of gravity in their X-ray burst simulations. As we argue in \S 4, the lightcurve probably does not depend too sensitively on gravity, but additional simulations are needed to check this, and to calculate the uncertainty in the predicted model flux which enters in equation (\ref{eq:fczcorr}) relating $f_c$ and $1+z$. On the other hand, for the comparison to spectral models, in Figure \ref{fig:mr}, the upper limits on $R_{\infty}$, represented by the solid colored curves, are placed in such a way as to coincide with the appropriate curve of constant surface gravity, consistent with the atmosphere spectral models used to derive those upper limits.

A second issue is that the upper limit on $R_\infty$ from the spectral models is based on fitting the initial part of the cooling tail only. We found that at first the slope of $K^{-1/4}$ with flux agrees well with the theoretical models of $f_c$. In the latter part of the burst, however, at lower fluxes, the agreement breaks down. For $F/F_{\rm Edd}\lesssim 0.2-0.3$, $K^{-1/4}$ increases with decreasing flux, but more rapidly than expected based on the predicted $f_c$ values, particularly those given by the solar metallicity models. Some other explanation is required for the rapid increase in $f_c$ and corresponding decrease in blackbody normalization in the tail of the burst. In 't Zand et al.~(2009) suggest that this decrease could be due to incorrect subtraction of the persistent emission, in particular, the subtraction of a thermal component that comes from the neutron star surface during accretion which is no longer present during the burst. Van Paradijs \& Lewin~(1986) pointed out that this effect should become important during the tail of the burst, when the burst flux becomes comparable to that of the accretion. Looking at Figure \ref{fig:fit}, the observations and the low metallicity (solar metallicity) models begin to deviate at fluxes below $\sim 5\times10^{-9}$ \fluxcgs ($\gtrsim 10^{-8}$ \fluxcgs) compared to the persistent flux of $2.1\times10^{-9}$ \fluxcgs. 

The disagreement between the observations and models begins sooner following the burst peak for solar metallicity than for low metallicity models because the former have a depression in $f_c$ at low fluxes $F/F_{\rm Edd}\lesssim 0.3$ (see Figure \ref{fig:fit}), arising from absorption edges in partially-ionized Fe (Suleimanov et al.~2011b). There is no sign of such a dip in the observations of \src. This suggests a low metallicity in the photosphere, contrasting with the conclusions of Galloway et al.~(2004) and Heger et al.~(2007) who argued that the metallicity was solar, based on burst lightcurves and energetics. A way to reconcile these disparate results is to consider the possibility that Fe, whose presence has a significant influence on the spectrum but not on the burst energetics, may be absent from the atmosphere during bursts. If accretion halts during the burst, then Fe will rapidly sink through the atmosphere (Bildsten, Chang, Paerels~2003). On the other hand, since bursts from \src\ are all sub-Eddington, accretion may continue during the burst, resupplying Fe to the photosphere. Furthermore, while disk accretion only deposits mass near the equator, the accreted mass spreads faster latitudinally ($\lesssim0.1\,$s; Inogamov \& Sunyaev 1999; Piro \& Bildsten 2007) than the timescale for Fe to sink through the atmosphere of $\sim1\,$s. Proton spallation could also destroy a substantial amount of the accreted Fe (Bildsten, Chang, Paerels~2003). It should be noted that X-ray bursts produce a wide range of elements in their ashes which could significantly alter the spectrum. However, mixing of burned material to the photosphere is thought not to occur due to the substantial entropy barrier (Joss 1977; Weinberg, Bildsten, \& Schatz 2006).

Another important unresolved issue is the changing spectral normalization $K$ with accretion rate. The blackbody normalization $K$ is $\approx 20$\% smaller for the 4.07 hr recurrence time bursts than the 5.74 hr recurrence time bursts.  We cannot explain this difference by changing the composition at the photosphere and therefore changing $f_c$ (see discussion in \S 5.3).  Also, it seems unlikely that a major change in composition would occur with only a $\approx 50$\% change in accretion rate and smaller change in burst energy and lightcurves. As mentioned previously in \S 6, Suleimanov et al.~(2011b) discuss large color correction factors associated with accretion onto the neutron star equator, which they suggest accounts for the variations in measured $K$ for the different spectral states of 4U~1724-307. Whether the 50\% increase of accretion rate seen  in GS 1826-24 could result in the amount of hardening of the spectrum observed needs to be investigated. If disk accretion onto the star significantly hardens the burst spectrum, it considerably complicates inference of the mass and radius from burst observations, and means that the range of $f_c$ included when calculating errors on mass radius determinations should allow for a larger range of values than given by spectral models. 

We have found that, even though they do not reach Eddington luminosity, the bursts from \src\ show enough dynamic range in flux as they cool to significantly constrain $F_{\rm Edd}$ by comparing with spectral models. A promising source to look at further is KS~1731-254 which shows both mixed H/He bursts with similar spectral evolution to \src\ (Galloway \& Lampe 2011) and photospheric radius expansion bursts. Analysis of these different bursts, which occur at different persistent fluxes and involve different fuel compositions (based on their energetics, peak luminosities and durations), would give a stringent test of the spectral models and help to constrain any additional spectral components.

\acknowledgements 
This work was supported by the National Sciences and Engineering Research Council of Canada (NSERC) and the Canadian Institute for Advanced Research (CIFAR). DKG is the recipient of an Australian Research Council Future Fellowship (project FT0991598). The authors are members of an International Team in Space Science on type I X-ray bursts sponsored by the International Space Science Institute (ISSI) in Bern, and we thank ISSI for hospitality during part of this work.

\end{document}